# A Critical Take on Privacy in a Datafied Society

Marco Cremonini

University of Milan, Italy

`marco.cremonini@unimi.it`

*Abstract*

Privacy is an increasingly feeble constituent of the present datafied world and apparently the reason for that is clear: powerful actors worked to invade everyone's privacy for commercial and surveillance purposes. The existence of those actors and their agendas is undeniable, but the explanation is overly simplistic and contributed to create a narrative that tends to preserve the status quo. In this essay, I analyze several facets of the lack of online privacy and idiosyncrasies exhibited by privacy advocates, together with characteristics of the industry mostly responsible for the datafication process and why its asserted high effectiveness should be openly inquired. Then I discuss of possible effects of datafication on human behavior, the prevalent market-oriented assumption at the base of online privacy, and some emerging adaptation strategies. In the last part, the regulatory approach to online privacy is considered. The EU's GDPR is praised as the reference case of modern privacy regulations, but the same success hinders critical aspects that also emerged, from the quirks of the institutional decision process, to the flaws of the informed consent principle. A glimpse on the likely problematic future is provided with a discussion on privacy related aspects of EU, UK, and China's proposed generative AI policies.

## The Quest for Online Privacy

It does not exist anything resembling to a general concept of *online privacy*, if it ever existed at all. At most, some online services preserve users' privacy, but only to a certain extent and often in a way not clearly verifiable. This is a bold statement, of course, and as almost all bold statements, it is logically weak because reality is always more nuanced than that. However, it is hard to sustain that it is fundamentally false. On the contrary, it has become popular to acknowledge the fact that whatever we do online, we are all constantly watched, monitored, profiled, targeted, tracked, recorded, and that this mass of data and metadata is somehow exchanged in remote data centers, among obscure agencies, companies, advertisers, businesses.

Nobody is any longer shocked about this, it has become common belief even among laypersons. Maybe we get just a little surprised when we discover that the actual extent of the absence of online privacy is even broader, deeper, and more systematic than we had believed. But it is a state of surprise that is quickly discounted and set as the new normal, the new homeostatic level of privacy absence [Slovic 1982, Wilde 2014]. As a matter of terminology, even talking of *privacy loss* or *privacy preserving* is largely a figure of speech rather than a real description, because they implicitly support the false narrative that something we had or we owned has been taken from us, a scam operated against us and someone run away with what was in our possession. It is the belief that there was a time when everything was fine—remember, you could have been a dog on the internet and nobody would have known? [Fleishman 2000]— then they came and stole that privacy from us. But this is just a false narrative, a process of myth construction of a golden era followed by a dark age. It does not reflect what really happened. Those days of alleged exceptional privacy were the early years of the internet when everything was different and most of all everything was small, incredibly tiny, if compared to today's digital world. Then the growth begun, exponentially, but since then privacy always had no place in the commercial, advertising-based, globalized internet, it was simply an inconvenient by-product of technical limitations, with the obstacles to a complete absence of online privacy that have been systematically removed. The natural state of the internet has always been of *privacy absence*; we should more appropriately talk about *privacy gain* to refer to those limited areas where some degree of privacy has been achieved, rather than mentioning *privacy loss*, which implies a misleading reference point. Similarly, for tools and techniques, the expression *privacy preserving* should be adjusted to reflect the fact that they typically add a feature to the normal state, rather than preserve a property that was never granted. Setting a common reference point is the basis for fruitful discussions and analyses, it is not worthless to rethink the narrative that guided the debate on privacy in the last decade.

The absence of online privacy has been normalized in public opinion. This is perhaps the most relevant privacy-related fact to register from the last decade. Not privacy-related research or technical advances, not privacy and data protection laws and regulations, not political initiatives, none of these have structurally changed the landscape. At most, limited and localized improvements have been achieved, with respect to the premises of a decade ago. But what strikes out is that we have not even figured out exactly what lays behind the infamous *privacy paradox*, i.e. why people declare to have privacy in great esteem, but act often as if they do not really care about it, a dilemma that has fueled so many lively debates [Acquisti 2009, Utz 2009, Xu 2011, Acquisti 2016, Barth 2017]. On the other hand, we are still reluctant to openly admit the evidence of the wrong premises, theoretical and practical, supporting many initiatives, solutions, laws and regulations, all centered around the flawed assumption of *informed consent*. At any rate, the *privacy paradox* and the *informed consent* have not aged well during the last decade, at the point of having lost most of their interest and traction, may be one is still a hypothetical paradox and the other a hypothetical consent, but just two mild troubles among the many that coexist in the background of our society. People have adapted to an environment where their digital life is continuously exposed, and they adopted behavioral strategies, different and subjective, because there is never just one adaptive strategy [Beaudry 2005, Miklós-Thal 2023]. Some privacy-preserving tools have become standard during the technical innovation process of these years, others have proved useless, impractical, or plainly absurd, some have demonstrated effective functionality and are available as viable alternatives for the most privacy conscious among the users. However, no tool or technical innovation ever resembled like a general effective solution for everybody. Not even close. The same could be said for laws and regulations, attempting to crystallize scenarios in perennial transition and adaptation.

This is why still referring to a concept generically called *online privacy* increasingly sounds as a misleading, to say the least, representation of a fictional reality, because it looks like we are referring to something real, partially real, or potentially real, when instead, it is probably an ill-conceived paraphrase brought to the digital realm of a concept and a legal right born in the 20[th] century under circumstances and with motivations completely different than the current ones. Differences that are ever increasing. Lessig, in his book *Free Culture* [Lessig 2004] provided an excellent description of the misleading notion of privacy as *friction*: "The highly inefficient architecture of real space means we all enjoy a fairly robust amount of privacy. That privacy is guaranteed to us by friction. Not by law [...] and in many places, not by norms [...] but instead, by the costs that friction imposes on anyone who would want to spy. [...] Enter the Internet, where the cost of tracking browsing in particular has become quite tiny. [...] The friction has disappeared, and hence any 'privacy' protected by the friction disappears, too." Lessig noted in 2004 this loss of friction that had previously provided a certain amount of online privacy. Two decades later, the best hope for online privacy is still a kind of friction, this time in the form of human behavior failing to follow the logic of the market-oriented model of privacy, as I will discuss in the following. Other than that, privacy online is still strictly considered an individual issue [e.g., Acquisti 2020], not a societal one, although the nature of privacy as a social good has been recognized and debated for decades [e.g., Margulis 2003], with tech entrepreneurs and digital rights advocates as the main supporters of the individualistic vision, a prevailing approach that probably is at the root of the absence of privacy online.

Recently, we had the opportunity to watch one of the most striking demonstrations of the deleterious clash between privacy as strictly an individual right and privacy as part of the societal welfare with the spectacular worldwide failure of apps for digital contact tracing developed during the Covid-19 pandemic, at least in Western countries [Butler 2022, Márquez 2022, Karanasios 2023]. All those apps suffered from a public perception of scarce usefulness mostly for the difficulties to integrate information they collected with established physical contact tracing procedures and epidemiological plans to contain the contagion [White 2021a, VoxEU 2021, Ussai 2022, Vogt 2022]. A specific factor common to all of them exacerbated those difficulties and ultimately made them worthless for any epidemiological study and containment plan: the role of the most vocal and ideological among privacy activists and technologists to instill in the public opinion the almost conspiratorial belief that digital contact tracing apps could have acted as trojan horses of the government for eroding privacy rights [Privacy International 2020a, Privacy International 2020b] or as discriminating factors for the weakest sectors of society [Davis 2020]. Based on these largely unsubstantiated claims and distorted threat assessments, with the support from interested parties in the tech industry (e.g., [Apple 2020, Sharon 2021]), and the obstinate insistence on theoretical privacy risks, rather than on the concrete effectiveness in containing the pandemic and understanding its evolution, both from the press (e.g., [Shevlane 2020]) and academia (e.g. [Ahmed 2020, Hern 2020]), those contact tracing apps implemented the most inflexible version of privacy as an inviolable individual right, regardless of any outcome on public welfare [Davies 2021, Rich 2021]. That happened despite a more nuanced approach was suggested even by leading security and privacy scholars [Anderson 2020] and experimental evidence against the claimed prominence of privacy concerns emerged since the initial weeks of digital contact tracing apps [White 2021b, Horvath 2022]. As a consequence, none of those apps provided even a marginal benefit to containment procedures, the data collected were useless for any epidemiological study and analysis of the diffusion of the contagion, the waste of taxpayers' money in all countries has been staggering, and the public opinion ended up even more twisted towards conspiratorial and misjudged belief regarding privacy.

But there is even more to note in the quest for privacy that has been pursued in the last decade, something that is similar to a secret conceived in plain sight. It is the discomforting (from a certain perspective, clearly) evidence that for all the pervasive data collection that is happening, all the surveillance and microtargeting, all the gigantic effort to turn the digital realm into a dragnet of personal information, the effectiveness of models aiming to predict human behaviors remains often tragically low and unreliable. This happens almost everywhere, in advertising, political campaigns, epidemiology, social media, and so forth [Neumann 2019]. May be it is because the science of modelling human behavior is still underdeveloped, may be the gigantic amount of personal data is not sufficiently vast, may be future AI techniques will revolutionize the art of predicting behavior, all these potential explanations are possibly true, but as a matter of fact, what we know for sure is that the prediction ability based on personal data has been so often over-hyped to have become another false narrative haunting the discourse around privacy. Advertising, in all its forms, from traditional commercial advertising to political influence campaigns, has typically escaped an open and objective scrutiny, remaining instead a secretive sector with regard to practices, metrics, and assessment procedures of the effectiveness of their techniques. At least publicly, the effectiveness of advertising has largely been based on anecdotal evidence, subjective claims of interested commentators, and carefully crafted storytelling that, for digital advertising, inevitably ends up assuming as obviously true the irresistible power of data-driven advertising in knowing exactly how people behave and how to nudge them in the prescribed direction. In the privacy camp, the opposite storytelling abounds too. We all have been abundantly fed with horror stories of advertisers, for example knowing about the pregnancy status of women before everybody else, or about the threat to democracy due to devious social media campaigns based on mysterious profiling techniques. The Cambridge Analytica scandal has been the epitome of all these horror stories, at the same time feeding on two opposite appetites, the fear for the dark power of algorithms manipulating human behavior and the admiration for that seemingly superhuman power. Very few, at least initially, had doubts about the spin that the press gave to the story, focusing exclusively on the violation of Facebook terms of service and lack of controls, which led to the illegal gathering of millions of user profiles. In other terms, the message was to consider the breach of individual privacy leading to manipulation of electors as the only news [Cadwalladr 2018]. With regard to the meaning and relevance of the data, nothing more than the stereotypical "data is the new oil" [Cao 2018]. However, that was only how the press choose to tell the story. What happened had characteristics of the most traditional type of scam based on selling a magic potion to gullible buyers, this time in the form of obscure data-driven algorithms and models designed by a small political consulting firm. The asserted power of the microtargeting practice called "psychographic marketing" employed by Cambridge Analytica was never established other than some claims by interested parties [OPF 2020]. As a matter of fact, at best its potential effectiveness is debated, ranging from a possible but unstable benefit [Matz 2017, 2018] to a negligible gain with respect to traditional marketing techniques [Sharp 2018], at worst it proved no better than "scant science" [Gibney 2018].

The narrative centered on the indisputable and absolute power of personal data is necessary to sustain the reasons of both current data-harvesting industries and the hardliners among privacy advocates. But it is a toxic narrative that does not stand on solid ground and should be openly debated and scrutinized. I began this introductory section with a bold statement, often repeated without the needed context and nuances, and terminate it referring to an equally bold assumption, that of the prodigious power of personal data, which has represented the main pillar for both their frantic harvesting aimed at profiling human behaviors and the often-dramatized scenarios depicted by privacy guardians. Debiasing those bold statements and assumptions through rigorous scrutiny should be among the main goals for future online privacy analyses and research.

*The Many Definitions of Privacy*

Looking at the origin of the privacy concept, Aristotle's distinction between the public sphere of politics and the private sphere of the family is often considered the root. The distinction between public and private sphere has been one of the most debated in modern political science and philosophy. Goldschmidt, for instance, a renowned political scientist of the past century, speaking about democratic values of Western societies, mentioned "two disturbing tendencies: first, a tendency toward too much publicity with a consequent disregard of the individual's right of privacy; and second, a tendency toward too little publicity, with a consequent increase of secrecy in areas hitherto considered public" [Goldschmidt 1954].

With respect to the social transformations brought by market forces eroding citizens' privacy, the German philosopher Jurgen Habermas, in one of his most influential works [Habermas 1991], an account of how the liberal public sphere took shape at the time of a developing market economy, wrote that "the family now evolved even more into a consumer of income and leisure time, into the recipient of publicly guaranteed compensations and support services. Private autonomy was maintained not so much in functions of control as in functions of consumption […]. As a result, there arose the illusion of an intensified privacy in an interior domain whose scope had shrunk to comprise the conjugal family only insofar as it constituted a community of consumers."

It is important for this analysis to highlight that, despite the radically changed context, the debate on the concept of privacy went on during the whole 20$^{th}$ century, for the shrinking domain of privacy on the one side and the tendency to over-secrecy on the other. The concept of privacy has always been enmeshed with the definition of public realm, the interests of the industrial sector, the state's monopoly of power, the extension and limitations of property rights, and with the rise of human rights movements. This complicate network of relationships still exists and determines how privacy is shaped in the digital realm, who influences the narrative that frames the debate, what inspires regulators, especially politicians, obviously very sensitive to the mainstream opinions, how technological industries design and promote their products, how they make attempts to create markets for privacy or, on the contrary, markets for data-driven services, how the press and media catch up with these trends, and finally how scientific community and academia are shaped, informed, and react to the changing technical and cultural landscape related to privacy online. In short, defining what is privacy and what is the right to privacy has never been easy, always riddled with contradictions and oversimplifications, stretched in one direction or another by interested parties [McCloskey 1980]. The modern digital environment, for all its ambiguities and lack of nuanced analyses, has worsened that historical trend [Mulligan 2016].

Daniel J. Solove put it bluntly in his *A Taxonomy of Privacy* [Solove 2006]: "Privacy seems to be about everything, and therefore it appears to be nothing." Privacy as a concept traverses many disciplines and areas of expertise, and for this reason has accumulated dozens of different definitions, often with no common elements, sometimes even conflicting. Several commentators have lamented the lack of coherence or of meaning of the concept, and this has probably contributed to a general inhomogeneity of privacy-related studies. Posner called the term privacy a *misnomer* [Posner 1978], a term that tries to catch too many meanings and for this misses its significance. Kohn in her book [Kohn 2004] introduces the notion of *cluster concept* or *term*, that is a term that has multiple and sometimes contradictory definitions, so it can be defined only by a list of criteria, possibly weighted, being no one of those necessary or sufficient. Privacy, like game, public space, or democracy are examples of cluster concepts.

Smith et. al. [Smith 2011] present a classification of *information* privacy research distinguished from *physical* privacy. It is a popular classification, very intuitive, indeed. Privacy *online* refers to information privacy, mostly in the form of personal data managed by online services or online data generated by individuals. However, information and physical privacy are not disconnected, violation of information privacy may lead to violations of personal privacy, when, for example, the gathering of personal information about the habits of an individual may allow physical surveillance. On the other hand, a physical privacy violation may provide the means to access personal data. But perhaps more relevant than those mostly unlawful cases, mutual dependency between the informational and the physical domains is already an essential part of the overall scenario, and the more the online and the offline life of individual are enmeshed, the more the distinction between information and physical privacy becomes blurred, possibly misleading. It is not necessary to introduce criminal actions to discuss about this mutual dependency. Advertising is totally legitimate and has solid reasons to exist, the same is for the advances and applications of neural network-based AI. Nevertheless, in both these cases, the informational and the physical domains have mutual reinforcements, personal information gathered from one domain have direct influence over the other one, so are changes in one domain that trigger adaptations in the other. Hence, the traditional distinction between physical and symbolic, online and offline, or real and virtual is losing its representativeness even for privacy.

For information privacy, Smith et al. introduce two broad categories: *value-based* and *cognate-based* [Smith 2011]. In the first case, privacy is seen as a human right belonging to the society's moral value system. This approach is close to the origins of privacy as an individual right and to the distinction between public and private sphere. The digital society has blurred the traditional lines between what is considered private and what is public, not just by expanding the domain of public information over private ones (think to the role of social networks, for example), but also *vice versa*, increasing even more the tendency to privatize large parts of what were once public spaces (the online equivalent of private malls replacing public squares could be the prevalence of corporate owned and regulated discussion board, with a common aphorism "There's no Central Park on the Web"). Legal scholars have widely adopted a value-based approach to information privacy. Economists did too, in general, but with the difference that privacy is regarded as a *commodity* that, in principle, could be traded [Bennett 2011, Angwin 2014].

With the cognate-based approach, instead, privacy is conceptualized as related to the individual's mind, perception and cognition. Psychologists and cognitive scientists have discussed privacy for decades, but for online privacy the most relevant conceptualization is to consider privacy as a problem of *control* of access to self. This view has inspired some of the most important policies for data protection and privacy preservation, those whose pillar is to provide individuals with instruments to control the access and usage of personal information. More details on privacy policies and informed consent will be discussed in a following section.

**On Market and Industry**

In an article published in The New York Times on June 16, 2012, the journalist Natasha Singer opened her piece with messianic tone: "It knows who you are. It knows where you live. It knows what you do" [Singer 2012]. This undisclosed "it", she continued, "peers deeper into American life than the F.B.I. or the I.R.S., or those prying digital eyes at Facebook and Google". "It" was the *data broker* industry, meaning those companies that collect information on citizens from all possible sources, digital and physical, categorize them into thousands of profiles and sell them (or license for specific uses) mostly to the advertising industry, always looking for data to feed personalized online ads to people [McKenna 2013]. Typical categories of data managed by data brokers regard retail purchases, web site visits, financial data, location data, and health data.

It is worth noting that citizens' personal data have been regularly traded. This is not a secret, neither the peculiar by-product of the internet-based economy. It was the 1970 when the US Federal Trade Commission (FTC) promulgated the Fair Credit Reporting Act (FCRA), which sought to regulate how consumer data must be used by consumer reporting agencies in decisions about credit, employment, insurance, housing, and the like. At that time, a market for the collection, trading, and usage of citizens' information was already established. However, the FCRA did not cover the trade of citizens' data for advertising or other goals different than financial credit. It was only in late 1990s that other business-oriented practices involving customers' data begun to be analyzed. Nowadays, the scenario has evolved and, as typically happens, it became more multifaceted, with new players, different relations and weights, new insights and more specific analyses. In short, it is no longer that straightforward, with data brokers (and intelligence agencies) as the only villains of the privacy saga. For instance, data brokers and advertisers were clearly distinct entities until the first decade of 2000s, but the roles progressively blurred with the fast growth of the digital advertising market, which in 2018 surpassed the non-digital one in spending and today accounts for nearly 70% of total ad spending [Statista 2023]. In the digital advertising market, the biggest players are Google, Amazon, Meta, Baidu and the other large internet companies, whose role is unconventional, acting as advertisers, although not in the traditional sense but as providers of ad networks/ad exchanges, while denying their role as data brokers because not explicitly involved in marketing personal data (some would disagree, though [Hoofnagle 2021]). At any rate, by leveraging their massive user bases, they have been able to revolutionize the advertising sector. The rise of those giants, however, does not mean that more traditional data brokers, those collecting and aggregating data but not directly acting also as advertisers, have disappeared. On the contrary, companies like Axciom, Epsilon, and Equifax among the many still thrive and are able to collect personal data about billions of people worldwide. The Data Brokerage research project at Duke University provides a rich documentation about this elusive multi-billion industrial sector [Data Brokerage 2023]. At the very least, the existence of this integrated and highly sophisticated ecosystem means that everybody, everywhere, which habitually accesses information and services online or uses services that are digitally managed (from retail purchases to medical diagnoses) should realistically assume that not just all his/her online searches and visits are categorized and used for building commercial profiles, possibly sold or may be used directly by a tech giant, but also most of his/her financial, professional, location, and health data are ingested, categorized, aggregated, and transformed in commercially valuable information that sustain the enormously wealthy data industry. The degree of privacy that one could realistically assume to have is none or, for who lives under the jurisdiction of modern privacy protection laws, very low, possibly sensibly less than theoretically granted.

Some other interesting figures from Statista [Statista 2023] (similar estimates are reported by different sources with some differences in quantification but qualitatively equivalent) show that *programmatic advertising* (i.e., targeted-based advertising, hence more dependent from users profiling) is ever increasing and represents more than 80% of digital advertising spending, which combined with the increasing total amount of advertising spending, suggests that the frenzy for ingesting personal data is increasing too. Another information which has some relation with privacy is that mobile advertising is now largely predominant with respect to desktop advertising and increasing, which is bad for privacy because traditionally mobile apps have looser controls on information and possess more valuable personal data (e.g., location, health data).

Let's enter the new main character of the data collection scene: major companies of the exploding neural network-based, generative AI market, all formally independent but backed by one or more tech giants, with Microsoft and Google as the largest financing partners. Based on one of the few comprehensive analyses of trends in machine learning training requirement [Sevilla 2022], the last large-scale models exhibit a staggering increase in size up to 5 orders of magnitude with respect to previous deep learning models and a doubling time of less than 10 months (i.e., the doubling time is the time period needed for the required training datasets to double in size). In short, this trend means that necessary training datasets are gigantic and fast-increasing in size. It comes with no surprise the renewed interest in producing reliable synthetic data for training such systems, at least as a complement to real data [Nikolenko 2021], although machine learning system trained on synthetic data are notoriously haunted by problems of difficult solution [Shumailov 2023]. Initial estimates of the training costs of machine learning and large-language models AI systems, are in the hundreds of millions of dollars for current systems like OpenAI GPT and Anthropic GPT, heading towards the billions of dollar scale in 2030 [Cottier 2023]. These costs are justified by both the increasing demand of computational power and of data, to be collected or to be licensed from original owners. It is evident that the trend of development of modern machine learning systems is clashing with privacy, and no realistic compromise is in sight. Some are hoping in the salvific power of synthetic data, but with the caveat "if researchers can find the right balance between accuracy and fakery" [Savage 2023], which is actually a very strong condition, with respect to the state-of-the-art. Others put their best bet in privacy legislations, assumed to be potentially able to strike a balance between AI and privacy requirements, but only if AI *accountability* and *fairness* will be established, again at the moment not much more than two hypothetical long-term developments [Mulligan 2019, Kerry 2023]. *Ethical AI* is another approach that has commanded attention especially in academic and policy circles, with privacy included among the ethical principles. However, for the interest that the topic may have in discussions and essays, it looks unlikely that it might be the changing factor, at least in the short-term [Mittelstadt 2019]. If the current options to find a balanced trade-off between privacy and AI requirements seem flimsy, reasons to foresee a clash between them abound. An anticipation has already been provided by the case of the Italian data-protection authority that in April 2023 blocked for a few weeks the access to ChatGPT in Italy over privacy concerns related to age verification of underage users [GPDP 2023]. While for this specific case the decision has been generally considered excessively rigid, the case should not be dismissed as simply the extravagant decision of a local authority possibly overstepping its role. On the contrary, it could be a glimpse on the conflicts to come between national and supra-national regulators and the tech industry. Of these conflicting positions there is already ample evidence in the new initial proposals for regulating AI that have recently appeared in the EU, UK, and China. In all cases, the proposals include strong commitments, most likely not compatible with current economic and technical development trends of the AI sector. I will discuss in more details about privacy regulations in the next section. To conclude this summary about AI and privacy, in opposition to those expressing seemingly ingenuous hopes of a smooth integration between the two, there are those all gloom and doom about the future of privacy in a world of pervasive AI-based services [Manheim 2019, Walsh 2023]. More generally, the interest over modern AI potential and the surprising advancements made public with recent generative AI systems have, on the one hand, produced excitement and hype, but on the other sparked predictions so pessimistic to have earned the moniker of *AI doomerism* [Roose 2023, Wang 2023], which clearly would include privacy too. Taking the stand for one of these clear-cut positions might be comprehensible but it is unjustified from an analytical standpoint, privacy is too much a multifaceted and controversial issue and our current digital and technological landscape is increasingly complex, if not confused, to be satisfactorily described with a simple formula.

*Datafication and user behaviors*

*Datafication* is the neologism coined for the process of describing the various aspects of an individual life through a set of data, subsequently used in digital services, markets, recommendation, chat bots, or decision-making processes. A reasonable objection to this term would be that in a way or another, "datafication" has always been done, being the need or desire to describe people's lives quantitatively very common and ancient, almost ingrained in the concept of society. So, why the need now for a fancy neologism, if not for mere dialectic futility? The scale is the reason. One lesson we have certainly learnt in the decades of data-centered technologies and services is that when data are concerned, size and scale matter to an extent that was never recognized before. It may matter at the point that just scaling up the size of data, for example of the training datasets of a generative AI, might produce improvements in features unimaginable before. So, the true meaning of the *datafication* neologism, and the actual reason for its introduction, relies on size and scale of data and of the aspects of human life described through data; the partially untold assumption is that if datafication is sufficiently broad and data about human lives sufficiently vast, then we will be able to describe every human behavior with precision and accuracy. Said with a certain emphasis, the ultimate goal of datafication is to conquer the knowledge of human behavior by turning everything into data. Clearly, the motivating assumption is often taken for granted but in reality, it is still largely unsupported by evidence.

Datafication and privacy are concepts with evident conflicting goals, and as long as the datafication process proceeds, privacy becomes increasingly feeble. However, there is something subtler to note, based on feedback loops present in a society subject to the datafication process. The key issue is that behaviors naturally change in response to environmental stimuli and the datafication is now happening in plain sight, is known even to the laypersons, is discussed, may be in a disinformed fashion, but what matters more is that it is perceived as a fact and has become an actual stimulus for individuals in their perception of privacy, which is triggering different reactions, with the consequence that the context is changing. This represents the classical definition of a dynamical system, whose resulting states are usually difficult to predict and model, possibly unstable and depending on several coevolving factors. In short, this is probably the most relevant and challenging aspect to analyze with regard to datafication, behaviors, and privacy. Old-school cyberneticians knew very well this sort of self-adaptive effects between technology and society. Only analysts and scholars paying particular attention to societal adaptations have caught the changes that some strata of the population (e.g., citizens, buyers, users, voters) are putting in place as a response to datafication and its more evident products like targeted advertising, algorithmic propaganda, and behavioral nudging. Considering privacy gains, as in the past, they are more likely obtained by adding friction to the mainstream process, and this is what some scholars have observed is happening.

With respect to the advertising sector, from years dominated by programmatic digital ads, with the consequence of an increasing impulse on personal data collection and users profiling, hence disregard of privacy, several analysts have noted some glitches, may be the signal of some stresses, if not already a shift or an inversion of the trend. On The Harvard Business Review, Moorman et al. [Moorman 2022] observed an interesting anomaly in the August 2021 and February 2022 data about advertising spending: the spending in traditional advertising reported an increase after many years of steady decrease in favor of digital advertising spending. The deduction that an inversion of the trend has begun is likely still an unsupported hypothesis rather than a fact based on firm evidence, but nevertheless those pauses in the decline of traditional advertising might signal a state of fatigue of digital advertising. The article explicitly questions the presumed high effectiveness of digital advertising on the basis of an increasing "frustration and negative brand association with digital advertising clutter that prevents them [the users] from reading an article, watching a video, or browsing a website" [Moorman 2022]. On the contrary, user's engagement with traditional advertising seems increasing. Other possible related factors mentioned by the article are the popularity of podcasts and the decline of third-party cookies. While the analysis of the causal effects of these potential factors on advertising spending is still largely speculative, they witness an ongoing transformation of the context, which often traditional analyses of privacy do not consider. The premises on which digital advertising is built are slowly but steadily changing, with consequences still unclear but possibly relevant. This changing context will impact on privacy too. The fatigue observed in online users dealing with the often clogged, intrusive, and annoying digital ads presence has direct relations with studies on user attention as the scarcest resource for online marketing [Perlberg 2016, Belanche 2019, Santoso 2022]. A possible gain in privacy could be a by-product of this shift, if it would actually happen, a reduced perception of effectiveness in programmatic advertising may reduce the value of personal data and user profiling, hence relieve the pressure on their massive collection. This deduction is also speculative, the opposite may happen as well: a perceived lower effectiveness might push an even more massive personal data collection in response, with the motivation that such larger dataset would guarantee the claimed effectiveness of digital advertising. Hence, the only firm point is that the overall context is in motion and changes in policies, strategies, or legislations would probably be necessary. Users fatigue and *ad avoidance* have not emerged for commercial advertising only, political advertising, heavily invested in microtargeting of potential voters, suffers the same fate. Stubenvoll et al. [Stubenvoll 2022] have recently documented several ad avoidance strategies that have been observed in users targeted by political advertising, like *cognitive avoidance* (i.e., just ignoring or skipping over any ad), *blocking behavior* (i.e., actively acting to hide ads through technical means), and *privacy-protecting behaviors* (i.e., acting for denying the collection of personal data). With respect to privacy, the authors of the study added the wise observation, supported by some experimental result [Jung 2017], that not all ad avoidance actions are explicitly aimed at improving user's privacy, hence it would be unmotivated to claim that privacy is always a causal factor for those ad avoidance actions. It could be, but just partially. Again, privacy is often the by-product of users adding friction in the process, rather than the primary goal of the adaptive strategy of choice.

Catherine Tucker of MIT is one of the leading researchers in the study of the data broker industry, algorithmic processing of personal data and decision making, behavioral adaptations to programmatic advertising, and consequences on privacy. Her work sheds a light on the development of datafication and effects on individual privacy carried out in the last decade. To set the scene, it could be useful to remind that in 2014 the US Federal Trade Commission released a public report urging initiatives to enhance transparency and accountability of data brokers, whose practices and methods were deliberately kept inaccessible to public scrutiny for the previous two decades [Federal Trade Commission, 2014]. In the years that followed up to today, several relevant legislative initiatives have been proposed and some have been produced data protection regulations, in EU specifically and some US states, but so far not at federal level. As mentioned before, in the meantime, the data broker industry has expanded and for the most part has retained a high degree of paucity, combined with the fast growth of digital advertising and the predominant role of a few tech companies, which again have proved to be all but transparent with respect to data management and privacy. Hence, it seems fair to conclude that the plea of the US Federal Trade Commission had not much effect on a trend that instead has become even stronger.

Something else however has slowly but steadily emerged: the datafication has structural limits that increase the degree of uncertainty regarding the effectiveness of programmatic advertising, algorithmic decision making, microtargeting, and ultimately human behavior predictability and manipulability. In short, the intrinsic complexity of societies and collective human behaviors are haunting the simplistic assumption at the base of datafication.

*Algorithmic exclusion and privacy*

*Algorithmic exclusion* is an unexpected outcome from the datafication perspective. Algorithmic exclusion refers to the case of people ignored by algorithmic processing, hence excluded from the results of the algorithm prediction or decision. The reason why this happen is because for these individuals there is no sufficiently available or reliable data, so they just don't exist or are too vaguely profiled from a datafication perspective. This typically is a consequence of unequal conditions in the access to digital services among people, being the data trail one leave online or offline still largely dependent on the economic status. The stated assumption was that such cases of underprivileged hence scarcely represented in data would represent a fringe minority in a population. On the contrary, from a recently analysis by Tucker [Tucker 2023], it appears that the number of people involved is relevant, not a fringe minority. This has direct effects on the presumed effectiveness of datafication, which might turn to be strongly skewed towards some strata of the population. Consequences are missing or inaccurate algorithmic predictions regarding those underrepresented people, which may result in wrong recommendations, biased service provision, of unfair commercial and financial offers. Other consequences, however, may occur, with one particularly relevant for the analysis of the effectiveness of data brokers and the invasion of privacy. Following the anecdotic evidence regarding a lower than expected (or supposedly expected, given the opaqueness of the data broker industry) frequency of existing and accurate predictions about individuals [Neumann 2019], further experiments and analyses revealed surprising results demonstrating both scant effectiveness in predicting basic individual characteristics (*missing predictions*), for example the sex, even when based on a large number and variety of data [Neumann 2021], and high inaccuracy (*incorrect predictions*), for instance to predict race/ethnicity, again even when available data are abundant [Kaplan 20121]. In general, it appears that missing and incorrect predictions are highly correlated with wealth, education and home ownership, and the bias is accentuated when predictions refer to background or demographic aspects, less so for people's interests [Tucker 2023]. These results cast a shadow on the pretense of high effectiveness in predicting human characteristics and behavior of the datafication approach and calls for more and deeper analyses. With respect to privacy, if confirmed, the new results likely mean that friction is again working in favor of a certain privacy gain as a side-effect of the seemingly reduced effectiveness. However, as opacity of methods and procedures of data brokers helped sustaining the unsupported assumption of high accuracy and completeness, it equally makes difficult to reliably evaluate the actual privacy gain derived from the apparent ineffectiveness that has been observed. Therefore, opacity works in favor of the status quo represented by the two polarized factions, those boldly praising the benefits of datafication and those ideologically declaring the primacy of individual privacy over everything else. Dismembering the opaqueness surrounding data brokers, digital advertisers, and tech giants in the management of personal data should be the highest priority for the years to come, if the goal is to pragmatically and objectively evaluate the effect of datafication on digital services quality and privacy. This goal should probably have the precedence over new legislative initiatives or technical solutions, because the outcome of that analysis should be instrumental to legislative initiatives, procedural measures, or the design of new technology. It is however likely that this scenario will not occur, being the ambiguous status quo of datafication and privacy too precious to many stakeholders.

*Digital hermits and fully datafied users*

Another interesting outcome has been observed and studied with regard to people behavior living in the current environment characterized by an exhibited and outstanding level of datafication: the rise of adaptation strategies based on the degree of exposure to datafication and its inference and prediction features one is willing to accept [Miklós-Thal 2023]. These are individual strategies indeed, but with a clear relation to society, to affective, cultural, and relational connections, not simply dependent on an isolated selfish feeling of personal privacy in the traditional sense of "to be left alone" [Warren 1890]. The degree of exposure to datafication one is available to accept depends on individual values, on one's own self-representation, and on the societal circles he/she mostly belongs to. It is the complex network of relations "between ego and others" would say sociologists, that has been largely neglected by privacy as strictly an individual value and right.

The most relevant phenomenon observed in [Miklós-Thal 2023] is a tendency towards polarization in the adaptive strategies to face datafication. An assumption common in past privacy analyses inspired by economic models, which more or less explicitly considered personal data as goods that the owner could price and trade on a marketplace, was of being able to categorize personal data on a fine-grained scale of "privacy value". That was the basis for a meaningful economics of privacy [Acquisti 2016]. It turned out from [Miklós-Thal 2023] that the reality might be different and the rational assumption of people pricing personal information based on a value scale might be often false. Rather, the behavior could be different, in particular it is not the privacy value of personal information the economic good, but the information about oneself that could be possibly inferred from a piece of personal data. In this sense, a personal data that considered alone might have a low value, for example the purchase of a common product like a best-selling book, an act shared by a multitude of other individuals, when combined with other personal data might reveal with good accuracy a personal characteristic considered highly sensitive. Said differently, it is not just the value of single personal data that matter, but the inferences possibly derived from all personal data a person might think would be collected and the highly uncertain subjective perception of how good organizations ingesting personal data are in inferring highly sensitive information about a person. It is a matter of perception and uncertainty about the potential damage, discomfort, or disutility derived from datafication, not a rationality driven economic assessment of values and prices of personal data. Under this different interpretation advanced by Miklós-Thal et al. [Miklós-Thal 2023], which depicts a considerably different scenario with respect the one taken for granted by the economic-oriented approach to privacy, what was previously considered the best approach to managing personal data, i.e. a careful assessment and meticulous management of what data to release and with what degree of visibility or accessibility, becomes the less likely and less realistic. Perception of uncertain threats from largely incomprehensible counterparts inevitably results in poorly informed deductions, guesstimated evaluations, and ultimately in looking for heuristics, not figuring out analytical decision models. In short, black or white solutions emerge, hence polarization between those that tend to minimize the possible consequences of disclosing all personal data (*fully datafied individuals*) and those that try to remove all their digital trails and aim at the digital invisibility (*digital hermits*). Under the new assumptions, both adaptive strategies have valid reasons to be chosen. If this scenario is actually the one we are headed to in the modern digital landscape, then also several considerations about privacy should change, many of past analyses that had shaped the technical and scientific debate, as well as informed policies and regulations, are now problematic, in part or completely. The analysis presented in [Miklós-Thal 2023] adds other relevant details, in particular regarding the dynamics between firms gathering personal data and people adapting their behavior. The more the data collected, in a general context of very low or no privacy, the better becomes initially the modelling and prediction ability of firms, although with some limitations possibly stronger than expected. This improved ability given by the enormous amount of data collected and some remarkable technological advancements, together with the extreme level of hype about the possibilities derived from the new technology (for example, AI techniques), and the general lack of systematic public scrutiny of processes and methods employed by those firms, all contribute to generate a state of confusion and elusiveness close to the famous third law formulated by the science fiction writer Arthur C. Clarke that "any sufficiently advanced technology is indistinguishable from magic" [Clarke 1968]. In this scenario, the hypothesis of rational management of privacy is losing ground and the one of more instinctive, gut-feeling behaviors gets more credits. Another consequence is that, at least theoretically, firms running massive data collection would be better off slowing the pace of that activity and returning to a stealthier mode of operation, in order to prevent the partition between fully datafied users and digital hermits. It is with the more economic-oriented and rational approach that the effectiveness of the data collection improves in the long term. What about privacy? In the partitioned scenario, fully datafied users have no privacy at all and digital hermits might experience a significant level of privacy, but probably most important, the digital hermits would introduce friction in the overall process, making data incomplete, fragmented, the population for which data are sufficiently vast to guarantee good accuracy in predictions insufficient to represent the whole. Friction is key for privacy. On the other hand, in the rational scenario, where personal data are priced and traded, in the short term a higher level of privacy, on average, is likely to be obtained, but would it be defensible in the long term or, because of the rationality of the economic process, would it be progressively eroded? Currently, this is an open question still without solid analyses to articulate a thoughtful answer. What we know for sure is that the rational scenario is the basis for almost everything has been done for data protection and privacy so far, from economic-oriented models, academic research, to laws and regulations, all of them have postulated the benefit of choosing vary specifically on as value-based scale what personal information to disclose, how and to whom, which has its logical base on assuming the existence of a rational process of risk assessment and decision. This assumption is necessary for approaching privacy as an individual rather than a societal matter. However, the body of recent studies that I have briefly summarized in this section is pointing to other possible explanations. A more complex, dynamical, and multifaceted scenario is emerging, questioning the traditional approach strictly focused on privacy as an individual matter.

**On Regulatory Frameworks**

The decade 2013-2023 has seen the rise of the legalistic approach to data protection and privacy, which became not just the most publicly debated privacy topic (the other one, cryptography for end-to-end communications and for data storage had generally a more technical audience), but succeeded in obtaining concrete achievements, starting from what has become the model and reference for all data protection regulations so far: the 2016, effective from May 2018, EU's General Data Protection Regulation (GDPR) [European Parliament and the Council 2016]. Laws and regulations aimed at protecting personal and sensitive data have been defined earlier too, starting from the 80s [OECD 1980] and more in the 90s [European Parliament and the Council 1995]. With the GDPR, the field of privacy and data protection legislation reached a corner stone entering in the era of modern privacy regulatory frameworks for its reach, the number and extent of its prescriptive rules and principles, the level of fines, the much commented "privacy by design" and "privacy by default" principles, the new formal roles and authorities, and for the first time the formal homogeneity over a supra-national organization like the EU. In practice, all following privacy and data protection initiatives have been described with respect to the GDPR.

It is not the aim of this section to recall the history of privacy legislation, which might go back at least one century, or to summarize GDPR's principles and characteristics. There is plenty of bibliographic resources for these topics, which have been presented and debated endless of times. Instead, the focus is on what happened after the GDPR became effective, how the newest proposals are trying to adapt to the transforming landscape, including the early material for the future revision of GDPR, and what critical aspects emerged or persisted in regulatory frameworks. In the second part of the section, I will focus on the concept of *informed consent*, the pillar of data protections regulations from the beginning in the 80s up to now with the GDPR and other modern regulatory approaches to privacy and data protection. It was necessary years ago, it is increasingly urgent today to critically analyze the foundational hypothesis of the informed consent and its intrinsic strictly individualistic perspective, because of its paramount importance for the whole approach on privacy.

Starting with the analysis of the GDPR after its successful application, in 2020 the EU Commission published some observations about the first two years [European Commission, 2020]. Rightly, it observed that the GDPR represents a model for other countries, mentioning Chile, South Korea, Brazil, Japan, Kenya, India, California, and Indonesia. Truly a composite mix of countries from all corners of the world. To mention California is somehow odd, however, being it the obvious outlier in the country list. California is at the same time one of the main industrial and economic region of the world, the cradle of most digital innovations of the century, but also the birthplace of those tech corporations that have expelled privacy from the internet realm and the epicenter of the cultural revolution that brought the massive datafication of today. And evidently California does not represent the whole USA, which up to now has not followed GDPR's steps nor proposed a federal privacy regulation. Being the EU Commission's commentary written in the midst of the first wave of Covid-19, it cannot avoid to reference the pandemic, but it sounded self-congratulatory and proved shortsighted in extolling the virtues of privacy protections measures implemented in "solutions aimed at monitoring and containing the spread of the virus, calibrating public policy countermeasures, assisting patients or implementing exit strategies" [European Commission, 2020]. As previously discussed, those solutions proved to be a failure from the beginning, they never delivered any of the outcomes they were designed for, and the highest priority given to privacy protection in digital contact tracing solutions was never unanimously praised as the best choice, on the contrary it was openly questioned, criticized, and warned as highly ineffective and ill-conceived by experts of epidemic containment and contact tracing, as well as by some highly respected security and privacy scholars. Following the observation on digital tracing app, the EU Commission commentary cites "[P]rotecting personal data is also instrumental in preventing the manipulation of citizens' choices, in particular via the micro-targeting of voters based on the unlawful processing of personal data" [European Commission, 2020], an evident implicit reference to the Cambridge Analytica scandal of 2014, which, however, in 2020 was already been dissected and analyzed in details, with the actual effectiveness of voters microtargeting questioned by many commentators. The EU Commission is the institution in charge of proposing and enforcing the EU policy, therefore the release of slightly pretentious statements about achieved results is not surprising, what is interesting in this opening of the commentary is to observe the same style not uncommon in privacy reports, often indifferent to opposite opinions and evidences at the point of sounding ideologic, ready to jump to conclusions even when the evidence is scant, and leaning towards excessive simplification. This approach certainly does not serve well for future challenges of privacy. In the remaining part, the commentary mentions a still not perfect cooperation and harmonization of rules and approaches, which is reasonable given the short period of time elapsed and the difficulty and novelty of the process. Next, the empowering of users is stressed as one of the main goals, which on the one side is certainly positive because before the GDPR the balance of power was disproportionally leaning towards the organizations collecting and processing data, with few to no possibility for the users to demand a fair and correct management of personal data. On the other hand, however, the explicit focus on empowering users with individual rights accentuates the individualistic approach to privacy, the central role of the informed consent, the market orientation praising competition among data processing organizations, and the overall assumption of rational economic agents. All these aspects are highly critical, largely studied, and rightly questioned. Finally, the commentary states that "[T]he GDPR, having been conceived in a technology neutral way, is based on principles, and is therefore designed to cover new technologies as they develop" [European Commission, 2020], which is not just an overconfident claim, but a naïve prediction about future technologies and a simplistic assumption on the neutrality of principles. As we are seeing today, the mesmerizing potential of modern AI technologies has already disproved that claim. This points to another critical problem regarding privacy: the more the legalistic approach becomes central to privacy protection, the more relevant is the role of legislators, institutions, political parties, and the bureaucracy, but none of them, today, appears truly in control of the increasing complexity, dynamicity, uncertainty, coevolving nature of digital technologies and privacy, and market pressure.

For all its undisputable successes, the GDPR has also received critical analyses from independent organizations, highlighting distortions and unaccounted effects. A short report by the Center for Strategic and International Studies (CSIS) [Heine 2021] has commented, among others, on the apparent disparity of GDPR fines issued in EU countries. Few countries issued the majority of fines, with Spain disproportionately proactive even with respect to the other most active countries, but often for minor violations and small amount of money. From Enforcement Tracker (https://www.enforcementtracker.com/), a web site recording information on GDPR fines, at July 2023, around 20% of fines are not larger than 1000€ (roughly 400 over 2000), another 20% are between 1000 and 3000€, the following 20% of fines are between 3000 and 10000€, which means that approximately 60% of fines are smaller than 10000€. The first fine in the millions scale was issued by Spain in 2021 against Equifax Iberica, the Spanish branch of the US corporation specialized in collecting information about individuals credit history. The Irish Data Protection Commission was subject to specific criticisms for its suspicious inaction with respect to some of the major US tech corporations, several of them having their EU headquarters in Ireland. This has instilled doubts in many observers and analysts regarding the effective independence from those internet giants. This apparently has changed starting from late 2021, then 2022 and 2023 when Ireland issued some of the largest fines, from 5.5 million up to 1.2 billion of euro, to Meta corporation and its controlled WhatsApp, the largest fine for unlawfully transferring personal data to the USA [DPC 2023]. Google too was repeatedly fined in the period 2019-2021, in this case from the French Commission nationale de l'informatique et des libertés, with fines from 50 to 90 million of euro, the largest for the violation of cookie informed consent procedure [CNIL 2022]. While GDPR violations and consequent legal actions and fines against internet tech corporations like Meta and Google (Amazon too received the second largest fine from Luxembourg, but the decision is still disputed and not much details have been released) were hardly a surprise, because they seem the direct consequence of a data-centric business model that has its origin in the early years of 2000s, something newer and relevant is surfacing from data on GDPR violations. It is the case of Clearview AI (https://www.clearview.ai/), a USA facial recognition company founded in 2017 and explicitly positioning itself in the growing sector of AI-centric companies. Clearview AI proposes its services for law enforcement, and for securing people, facilities, and commerce. It is not this essay the appropriate place to analyze the validity of Clearview AI business proposition and the actual effectiveness of facial recognition technology. The most critical aspect, which is representative of a new whole class of problems between datafication and privacy, potentially with enormous implications, is the staggering amount of data needed to train these new AI technologies and models, being them for facial recognition or generative AI. Clearview AI indiscriminately scraped billions of images from the web, social networks, and online photo repositories [Kashmir 2020]. It was initially fined by Germany in 2020 for a minor violation of GDPR, but it was in 2022 that fines quickly escalated into the millions of euro and more issued by Italy, Greece, UK, and France (twice). Legislative initiatives escalated too, with the majority of members of the EU Parliament that voted for a total ban of real-time biometrics in public spaces to be included in a draft legislation managing AI applications called the *Artificial Intelligence Act* [Liboreiro 2023]. I will briefly discuss about emerging AI managing legislative initiatives in the following, for now it is worth noting that they still are in the draft stage (all except one) and that such a drastic measure as the total ban was on the one side contrasted by calls for a more nuanced approach admitting exceptions, in particular for law enforcement [Amendment 800 2023], and on the other of being too soft on privacy protection [Amnesty Intl 2023]. It is already well-established that the management of AI-centric services will be a legislative battle field between datafication expansion and privacy protection, with unpredictable consequences for industry and society.

The GDPR's first major review is due by May 2024 [McQuaid 2023, Goujard 2023] and the amendments will probably improve enforcing and cooperation, although the European Data Protection Board in its programme for 2023-2024 has a quite rich list of items [EDPB 2023]. A possible risk looming over the GDPR is represented by its own success, which might obfuscate the need for inevitable modifications to deal with a changing landscape. From the case of EU's GDPR, it appears that who of what (e.g., panels of selected experts, public consultations, industrial lobbyists, political parties, economic advisors, etc.) will orient the understanding and the approach of the EU Commission and Parliament is one of the most critical aspects to establish the future path of datafication and privacy.

*AI regulatory initiatives and implications on online privacy*
A much-anticipated legislative initiative by the EU is the future *Artificial Intelligence Act*, a comprehensive legislation of AI systems used in the EU, which will supposedly conform to principles of safety, transparency, traceability, non-discrimination, and environmental sustainability. In addition, those AI systems will require a human supervision, being the fully automated supervision assumed to be unable to prevent harmful decisions [European Parliament 2023]. Whether it is realistically possible to satisfy this list of required principles is doubtful. Similar legislative initiatives focused on AI have been suggested by others as a necessary measure both to exploit the vast economic potential of AI and to curb the potential (very likely, not just potential, according to a certain line of thought) degeneration or misuse of AI-systems into harmful, unfair, biased, dangerous, and even out-of-control autonomous systems. How to strike a balance between economic potential and threat mitigation is part of the geopolitical strategy of each regional power, with each one claiming to be on the verge to become the next super-power, for one reason or the other, of the AI-centric world. The EU, fortified by the success of the GDPR but weak on the industrial innovation side, is positioning itself as the next leader of the world with AI first respectful of human primacy and principles, then business friendly. Differently, the UK government (and likely the USA as well) seems more inclined to put business friendliness first, with ample emphasis on the imminent fourth industrial revolution, a not-so subtle reminder of the historical role of UK in the first industrial revolution [Donelan 2023]. The USA still seems to be at the early stage of press announcements from political leaders, supporting the usual friendly stance towards corporations [Lima 2023]. Both the EU and UK, even in their differences, explicitly mention privacy of personal data among the aspect threatened by the expansion of AI systems. Interestingly, a recent report by the UK Centre for Data Ethics and Innovation (CDEI) [CDEI 2023] showed a majority of people convinced that AI benefits will outweigh the risks, but the respondents also admitted to have limited awareness of AI, and that the invasion of privacy is one of the main risks. The public opinion seems confused, to say the least, and legislators risk rushing to establish laws and regulations without sufficient clue.

One country anticipated all others, but its case is hardly reported in the press or mentioned by public bodies in Europe or the USA: China. It often produces a mix of surprise and condescendence in Western commentators to know that China has proposed a public consultation with companies on a draft legislation called *Measures on Generative AI* [Daum 2023a], then issued the *Interim Measures for the Management of Generative Artificial Intelligence Services* [China Law Translate 2023]. Obviously, the Chinese state organization is very different from Western democracies, the role of the government and of the ruling party on Chinese economy and society has no parallel in the West. This is true also for privacy, for which citizens are assumed to have none with respect to the government. However, the vast differences should not be used to deny the existence of similarities. It would be a false claim to state that in China the right to individual privacy does not exist. As well as claiming that market competition, property rights, discrimination and misinformation prevention do not exist. They all exist and are regulated, even individual privacy, not with respect to the government, but definitely regulated and enforced with respect to everybody else, including other individuals and companies. Therefore, it is not useless to consider how China is dealing with the issue of regulating generative AI, in particular for what concern privacy and data protection. The draft legislation for privacy and data collection was stating that all users of generative AI services should have their identity verified, their personal data be collected only upon an informed consent and following the principle of minimum necessity. More specifically, the draft specified that where generative AI services involve personal information, the service providers are considered 'personal information handlers', which means that consent is required for the use of personal information in any training data. This was a particularly hard requirement, the same that when theorized for Western regulations could overheat the debate and produce the toughest opposition from different parties, the industry for once. Also, the Chinese draft was requiring that information given by users should not be used to create user profiles or be provided to other parties. In addition, if users' identities might be inferred, that input information cannot be recorded. On the other hand, service providers are required to monitor user input for illegal and negative information, reporting to authorities the violations and acting on the provision of service. In summary, given all the differences, China proposed a strict privacy regulation. That was the draft legislation open to comments from the industry, clearly interested in reducing the strictness of the prescriptions, and it succeeded, as is common and expected in this kind of negotiations. The measure actually enforced, in fact, contains a new article stating that industry's main interests in AI innovation should be weighed equally as security and governance. Specifically, exceptions are introduced by limiting the scope of the rules only to public-facing generative AI services. Private-facing and research services are exempted from the regulation. This clearly reflects the importance given to maintain competitiveness on a global scale. For public-facing generative AI services the prescriptions have been mostly maintained intact, including those regarding individual privacy, and even improved, clarifying that individuals have the right to request to access, reproduce, modify, supplement, or delete their own personal information [Daum 2023b]. The case of China is illustrative of the tension that everywhere exists between the datafication process and individual privacy, which will characterize the government of modern AI. Apparently, China has chosen a collaborative approach, initially stating the foundational principles but then demonstrating the willingness to support technical innovation and industrial competitiveness. China has the advantage over Western democracies to be free to decide fast and without the need to mediate among many stakeholders. It also has all the disadvantages that come with not being a democracy. For Western countries the decision process is certainly more complicated, for the EU even more so with inevitable pressures coming from national governments and a traditional uncertain attitude on innovation and development. The challenges ahead for regulating AI while protecting individual rights are enormous for everybody, and polarized positions sustained by certain advocacy groups or economic interests are probably the ones to avoid, in favor of more nuanced ones. For sure, a big unknown is how globally the main economic regions will behave if different approaches will be adopted. Arbitrage possibilities could arise, shift of power, different development trajectories and so forth are elements that could play important roles.

*Informed consent, dark patterns, and the illusionary control*

I conclude this section with a short summary of the most critical aspects concerning the informed consent requirement, probably the main pillar of all current data protection and privacy regulatory frameworks, including the most recent one, like proposed regulations for AI. In all cases, the concept of individual privacy and its enforcement through regulations stand on the assumption that each person has certain rights with respect to his/her personal data and he/she is entitled to decide, based on complete information, what, how, and when specific personal data should be disclosed to or collected by others. This in summary is the fundamental principle of informed consent: every person has to be asked if he/she is willing to accept to lose part of his/her privacy in exchange of something (a service, a good, information, etc.) and he/she has to know exactly what he/she is doing. The theory of the informed consent might be perfectly sound, the practice has always been riddled with many severe and apparently unsolvable problems, a reality that is perfectly known to almost everybody from a long time, way before the adoption of the informed consent for privacy.

The informed consent concept has an history that started about at the beginning of the '900 in medicine, with some legal decisions that established the principle of *patient autonomy*, which requires the consent from the patient to undergo surgery or other medical treatments [Beauchamp 2011, Bazzano 2021]. From those early landmark cases, the informed consent was extended to medical research leading much later to the birth of the bioethics. It was only in 1979 that in the USA, the Belmont Report officially stated the specific concepts of *information*, *comprehension*, and *voluntariness* as critical to the process of informed consent in medical research. However, it did not take long, just a few years, for problems to emerge. The uncertain meaning of the key concepts when translated into actual informed consent forms to be signed appeared immediately as crippling the meaningfulness of the informed consent concept [Bazzano 2021]. When does a patient could be said to be truly informed and has actually understood the full extent and consequences of the decision? Is the decision truly voluntary if alternatives are not presented thoroughly, objectively, and have been fully understood? These are some of the questions that have been endlessly repeated, and never had completely satisfying answers. A certain ambiguity and degree of freedom in interpreting the principles were always present and problematic. For example, some institutions exploited that to customize informed consents to the characteristics of particular research studies or to write the text to highlight some aspects of the research and minimizing others, or to try to sound captivating for the prospective participants. On the other hand, a certain flexibility was necessary to account for the range of researches involving human subjects [Bazzano 2021].

Today, in the digital realm, we would say that organizations devised *dark patterns* to lure people to accept the proposition [Mathur 2021], but for how fancy the expression "dark pattern" could be, it means not much more than trivial tricks, stylistic or dialectic most of the time, are employed to make people believe that they have decided in full autonomy and information, or to make it appear, from a legal standpoint, that informed consent has been genuinely expressed when instead it was not. The point, when we discuss about informed consent for privacy, is that there is basically nothing new in that, neither the concept nor the problems [Millum 2021, Laurijssen 2022].

Problems with the informed consent are all perfectly known from decades, and they were as such when privacy regulation frameworks were established for the first time in the '90s, when a modern framework like the GDPR was approved in 2016, and when new frameworks are proposed today [Gray 2021], all based on a literal and simplistic interpretation of the concept of informed consent. In all cases, regulatory frameworks have been designed knowing those fundamental flaws of informed consent and its troubled history, which in the biomedical field led to the development of the modern bioethics. The digital field has tried to mimic what has been done in bioethics but only superficially and in a largely ineffective way, approaching it mostly procedurally and through self-assessments. For example, the EU guidance for funding opportunities presents ethics and data protection requirements as coincident with the application of GDPR prescriptions [Hayes 2021], which states in generic terms the principles of informed consent and suffers from its well-known problems. This approach has inherited almost nothing from the bioethics experience and how it tried to mitigate the flaws of informed consent. For data protection, instead, ethics guidance reinforces and formalizes the ambiguity of the informed consent, rather than mitigating it. What appears is that the privacy and data protection field has never developed anything comparable to the analyses, debates, scrutiny, and ultimately experience and understanding accumulated in the biomedical sector, nor has truly made useful usage of that experience [Manson 2007].

Probably there is no better example of the failure of the informed consent assumption for data protection than the so called *Cookie Law* of the EU, first introduced by the 2002 *ePrivacy* directive [European Parliament 2002], amended in 2009 [European Parliament 2009], and later implicitly confirmed by the GDPR which defines cookies as user identifiers (i.e., Recital 30), making them personal data subjected to the informed consent for their usage, with the exclusion of few exceptions. These pieces of legislation are responsible for the decade and more of nonsensical proliferation of cookie consent pop-ups that since then have uselessly annoyed hundreds of millions of users and defied any reasonable meaning of informed consent, data protection, and privacy [Burgess 2021]. Such cookie-related pop-ups are often discussed as an example of dark patterns, and in several occasions, changes have been proposed to stop service providers trying to trick users with ridiculous subterfuges like camouflaging the rejection option with minuscule font size and light grey typeface or hiding it behind scornful panels of fine-grained configurations [Privacy International 2019]. Such trivial tactics are not the main problem, though. Even the plain unlawful collection of cookies, which is not uncommon, is not the main problem of this approach [Gikay 2022]. The problem is at the root of the approach and in the regulators, in the known flawed assumption that has been adopted without any consideration for the consequences, again well-documented in the biomedical sector, and most of all without any apparent ability to self-correct a plain wrong and nonsensical decision. The deluge of cookie request for consent is still plaguing the web, the European Commission has conducted a study in 2017 for the future revision of the *ePrivacy* directive and a result was in 2021 to state the obvious, i.e. "the cookie provision, which has resulted in an overload of consent requests for internet users, will be streamlined" [European Commission 2021]. Nothing concrete has been done so far, the paucity of the institutional decision process is high, as usual, with the only information that are unofficial and unconfirmed (e.g., [Härting 2022]).

The informed consent is and seemingly will be still for long the weakest spot of regulatory frameworks for individual privacy and data protection, and one important reason for the persistent state of absence of privacy online.

**Conclusion**

The debate on online privacy has been for too long and too often being stuck on explanations for the reason of its poor state largely influenced by analysts of different backgrounds (e.g., technologists, entrepreneurs, policy makers, law scholars, ethicists, etc.) giving full credits to polarized interpretations of the matter. On the one side those that at various degrees minimizes the relevance of privacy and the extent of its demise in favor of the hype over the extraordinary opportunities promised by data-centric solutions. On the other the doomerists, always backing seemingly conspiratorial scenarios of pervasive Orwellian-Huxleyan surveillance, and people transformed in puppets by the devious power of datafication. The third faction is represented by the supporters of the legalistic solution and its salvific properties, the only able to establish order from the chaotic situation created by technologists, the one able to make human principles to flourish.

Clearly, these are just caricatures of the different approaches, but it is true that these different approaches have systematically spoken without truly confronting with an opposition, different criteria of analyses and interests, and most importantly standing on claims rarely subject of open and accurate scrutiny. This state of affairs is deeply unsatisfactory to many.

This essay is an attempt to consider together several of the many faces of online privacy in a world, under the present conditions, that is inevitably headed to more datafication, more data-centric applications, and expanding related economic importance. Most of all, this essay highlights that in a mutable global scenario where pressures from technological innovations, geopolitics, and the industry are adding up, an online privacy traditionally highly influenced by ideologic opinion makers and silo mentality is hopeless. However, new insights and analyses from particularly attentive observers are telling us that there are opportunities for a revised approach to online privacy, mostly based on the need for systematic and scientifically sound investigations of the many undemonstrated assumptions and claims of the actors of datafication.

Beaudry, A. and Pinsonneault, A., 2005. Understanding user responses to information technology: A coping model of user adaptation. MIS quarterly, pp.493-524.

Belanche, D., 2019. Ethical limits to the intrusiveness of online advertising formats: a critical review of better ads standards, Journal of Marketing Communications, Vol. 25 No. 7, pp. 685-701, doi: 10.1080/13527266.2018.1562485.

Bennett, C. J., 2011. In defence of privacy: The concept and the regime. Surveillance & Society 8.4, 485.

Burgess, M., 2021, May. How to bypass and block infuriating cookie popups, Wired UK. Available at: https://www.wired.co.uk/article/cookie-popup-blocker-gdpr

Butler, M., 2022. Failed COVIDSafe app deleted, Department of Health and Aged Care, Commonwealth of Australia. Available at: https://www.health.gov.au/ministers/the-hon-mark-butler-mp/media/failed-covidsafe-app-deleted

Cadwalladr, C. and Graham-Harrison, E., 2018, March. Revealed: 50 million Facebook profiles harvested for Cambridge Analytica in major data breach, The Guardian. Available at: https://www.theguardian.com/news/2018/mar/17/cambridge-analytica-facebook-influence-us-election

Cao, S., 2018, April. Ex-Cambridge Analytica Director Unveils the Dark Reality of Data Industry, Observer. Available at: https://observer.com/2018/04/ex-cambridge-analytica-director-speaks-out-on-facebook-scandal/

CDEI, 2022, April. AI Governance, Full Report, Centre for Data Ethics and Innovation (CDEI). Available at: https://assets.publishing.service.gov.uk/government/uploads/system/uploads/attachment_data/file/1146010/CDEI_AI_White_Paper_Final_report.pdf

China Law Translate, 2023, July. Interim Measures for the Management of Generative Artificial Intelligence Services,. Available at: https://https://www.chinalawtranslate.com/en/generative-ai-interim/

Clarke, A.C., 1968, January. Clarke's Third Law on UFO's. Science. 159 (3812): 255.

CNIL, 2022, January. Cookies: la CNIL sanctionne GOOGLE à hauteur de 150 millions d'euros, Commission nationale de l'informatique et des libertés (CNIL). Available at: https://www.cnil.fr/fr/cookies-la-cnil-sanctionne-google-hauteur-de-150-millions-deuros

CNIL2022, October. Facial recognition: 20 million euros penalty against CLEARVIEW AI, Commission nationale de l'informatique et des libertés (CNIL). Available at: https://www.cnil.fr/en/facial-recognition-20-million-euros-penalty-against-clearview-ai

Cottier, B., 2023. Trends in the dollar training cost of machine learning systems, Epoch. Available at: https://epochai.org/blog/trends-in-the-dollar-training-cost-of-machine-learning-systems

Data Brokerage, Sanford School of Public Policy, Duke University, 2023. Available at: https://techpolicy.sanford.duke.edu/data-brokerage/

Daum, J., 2023, April. Measures on Generative AI, China Law Translate. Available at: https://www.chinalawtranslate.com/en/overview-of-draft-measures-on-generative-ai/

Daum, J., 2023, July. Key Changes to Generative AI Measures, China Law Translate. Available at: https://www.chinalawtranslate.com/en/key-changes-to-generative-ai-measures/
32